# Do Two-Level Systems and Boson Peak persist or vanish in hyperaged geological glasses of amber?

T. PÉREZ-CASTAÑEDA[+], R. J. JIMÉNEZ-RIOBÓO[♣] and M. A. RAMOS[+,*]

[+] Laboratorio de Bajas Temperaturas, Departamento de Física de la Materia Condensada, Condensed Matter Physics Center (IFIMAC) and Instituto de Ciencia de Materiales "Nicolás Cabrera", Universidad Autónoma de Madrid, Cantoblanco, E-28049 Madrid, Spain

[♣] Instituto de Ciencia de Materiales de Madrid (ICMM-CSIC), Cantoblanco, E-28049 Madrid, Spain

[*] Corresponding author: Email: miguel.ramos@uam.es

**Abstract**

In this work we extend, review and jointly discuss earlier experiments conducted by us in hyperaged geological glasses, either in Dominican amber (20 million years old) or in Spanish amber from El Soplao (110 million years old). After characterization of their thermodynamic and elastic properties (using Differential Scanning Calorimetry around the glass-transition temperature, and measuring mass density and sound velocity), their specific heat was measured at low and very low temperatures. By directly comparing pristine amber samples (i.e. highly stabilized polymer glasses after aging for millions of years) to the same samples after being totally or partially rejuvenated, we have found that the two most prominent universal "anomalous" low-temperature properties of glasses, namely the tunnelling two-level systems and the so-called "boson peak", persist essentially unchanged in both types of hyperaged geological glasses. Therefore, non-Debye low-energy excitations of glasses appear to be robust, intrinsic properties of non-crystalline solids which do not vanish by accessing to very deep states in the potential energy landscape.

## 1 Introduction

More than 40 years ago, Zeller and Pohl demonstrated beyond doubt that the main thermal properties of glasses did not follow at all the expected Debye behaviour as non-metallic crystals do [1, 2]. On the contrary, glasses and other amorphous solids, and even some disordered crystalline solids such as *orientational glasses* [3] and *glassy crystals* [4, 5], are known to exhibit thermal, dielectric and acoustic properties at low temperatures markedly different from those found in conventional crystalline solids [1, 2]. Furthermore, these low-temperature glassy properties show a remarkable degree of universality for any kind of glass, hence the term *universal* "glassy behaviour". Specifically, the specific heat of glasses depends quasilinearly on temperature, $C_p \propto T$, below about 1 K, followed by a broad maximum in $C_p/T^3$ at typically 3−10 K nowadays known as the "boson peak", that arises from a noteworthy and controversial broad peak in the Debye-reduced vibrational density of states $g(\omega)/\omega^2$ [2]. On the other hand, the thermal conductivity goes almost quadratically with temperature below 1 K, $\kappa \propto T^2$. In the boson-peak temperature range, the thermal conductivity of glasses exhibits a universal plateau [1, 2], in striking contrast with the behaviour observed in their crystalline counterparts. Both low-temperature thermal properties are therefore at odds with the cubic dependences predicted by Debye theory and successfully found for crystals at low enough temperatures [6].

The thermal properties of glasses below 1 K mentioned above, together with related acoustic and dielectric properties also at low temperatures [2], were soon well accounted for by the popular Tunnelling Model (TM) [7, 8], though some open questions remain unsolved [9, 10]. The central idea of the TM is the ubiquitous existence of small groups of atoms in amorphous solids due to the intrinsic atomic disorder, which can perform quantum tunnelling between two configurations of very similar energy, usually called tunnelling states or two-level systems (TLS). Interestingly, the ultimate origin of the corresponding thermal properties of glasses above 1 K and their low-frequency vibrational spectrum dominated by the boson peak at ~1 THz, are even more controversial and matter of vivid debate [11].

Indeed, the "anomalous" (in the sense of unexpectedly opposite to that of crystalline solids) universal behaviour of glasses at low temperatures is only one of the several *puzzles* pervading the scientific understanding of glasses [12]. The core of the problem is that the glass transition is a complex combination of kinetics and thermodynamics. We will not even mention the many different theories existing about the glass transition, but we would rather focus on the paradigm of the potential energy landscape (PEL) which is widely used. The PEL is basically a topographic view of the (3$N$+1) potential-energy hypersurface of any glass-forming substance [13–15]. It is used to be schematically projected on two dimensions for convenience, and is plenty of local minima and saddle points for thermal energies below that of the melting point for the stable crystalline state (the absolute minimum) when it exists (see Fig. 1). When a liquid is supercooled bypassing the crystallization down to the glass-transition temperature $T_g$, it becomes a glass getting trapped in one of the many possible local minima or metastable states, depending on the thermal history followed. Furthermore, many authors have speculated on the possible existence of an "ideal glass" which should correspond to the best and most stable possible glass achievable, associated with the lowest relative minimum. This ideal glass would have zero configurational entropy, equal to that of crystals, and has been associated to the possible existence of an underlying thermodynamic glass transition occurring at the so-called Kauzmann temperature $T_K$ [16].

Our main goal here has been to search out whether those abovementioned glassy excitations at low temperatures are robust and intrinsic properties of any non-crystalline solid, or rather they will gradually vanish when one obtains more and more stable glasses of lower and lower enthalpy or entropy, close to the ideal glass state, so drastically reducing the configurational disorder. Of course, many attempts of studying this issue can be found in the literature [17–23], but only through the very modest annealing or aging processes that could be conducted in a laboratory. Depending on the glass studied and several other factors, these works have given contradictory and inconclusive results. The answer is thus not clear, though most authors would likely envisage that if a drastic reduction of the frozen-in configurational disorder could be conducted, these glassy anomalies should be suppressed or, at least, significantly decreased. This is exactly what we wanted to realize, by trying to get very deep in the energy landscape, approaching the ideal glass state.

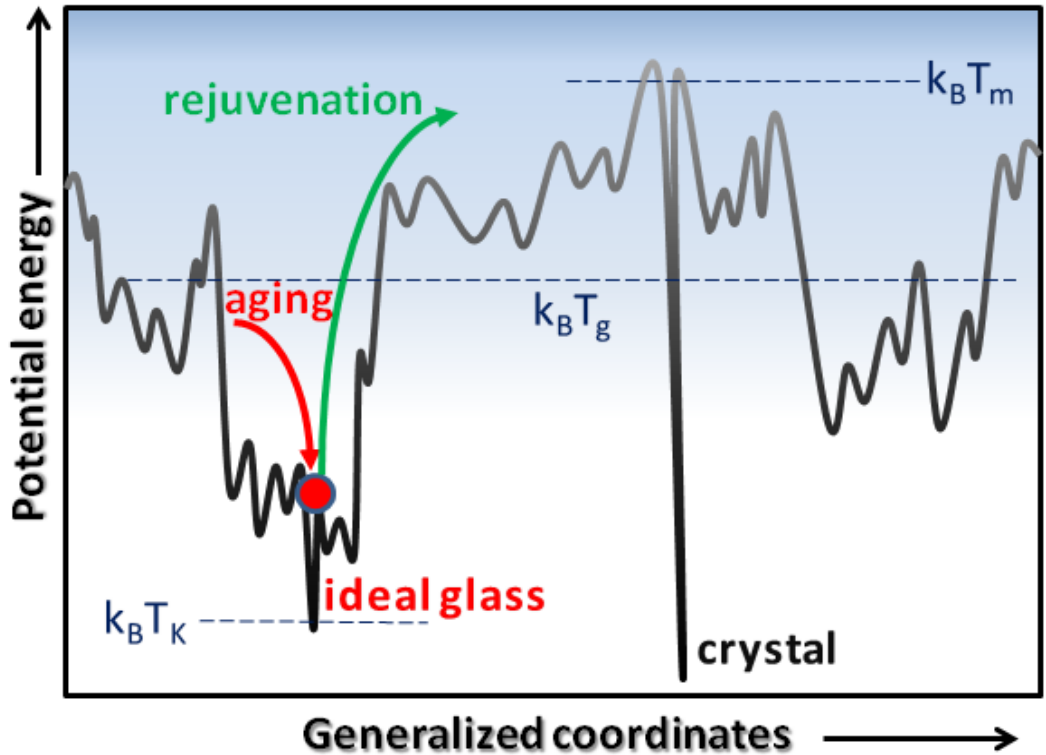

**Figure 1**. Schematic potential energy landscape (PEL) for supercooled liquids and glasses, including the hypothetical absolute minimum for a crystal state (not present in the case of amber). An ideal glass would be obtained after an infinitely long aging at the Kauzmann temperature $T_K$.

With such an aim, we have identified and studied a material which has been spontaneously subjected to unusually strong processes of thermodynamic and kinetic stabilization: amber, the well-known geological glass. Amber is essentially a fossilized tree resin that polymerized millions of years ago. Therefore, amber is a unique example of a (polymer) glass that has aged far longer than any system accessible in the laboratory, thus reaching a state (of lower enthalpy and entropy) which is not accessible under normal experimental conditions. In brief, it is an amorphous solid or glass which has experienced an extreme thermodynamic stabilization process (*hyperaging*) [24, 25].

In the next section, basic experimental details are given about the samples studied and the techniques and methods employed. Then, the main experimental results are presented, including sound velocity measurements, Differential Scanning Calorimetry (DSC) around the glass transition, and specific heat measurements at low temperatures. These results are discussed in the following section. Finally, a Conclusion section summarizes our findings.

## 2 Experimental details

### *2.1 Amber samples*

We have studied amber samples from two different sources. First, we have studied Dominican amber samples, about 20 million years old. By using standard chemical analysis, the composition of the samples was found to be 78.6% C, 9.9% H and 11% O in mass, in reasonable agreement with other values found in the literature [26], and with a nitrogen content ~0.1% and sulphur impurities below 0.2%. Second, we have studied amber samples obtained from a new deposit discovered quite recently in Cantabria, a northern region of Spain, within El Soplao territory, which have been dated to be 110-112 million years old [27]. Although our most exhaustive and detailed experiments have been conducted with the latter Spanish amber [25], it is also interesting to show here and concurrently discuss our earlier experiments on Dominican amber [24], as a complementary check of reproducibility and generality of the found phenomena. As a matter of fact, McKenna and coworkers [28] have recently studied the glass transition of amber from different sources and ages, exhibiting a wide variation of $T_g$'s with no correlation with amber's age, further finding that many of them were unstable and showed chemical reactions during the measurement. However, some kind of Dominican ambers and the Spanish amber from El Soplao that we used, were among the few stable ambers.

Samples for DSC experiments were prepared by manually milling the amber pieces in an agate mortar until a homogeneous particle size of some tens of microns was achieved. For the low-temperature specific-heat measurements, amber was cut into flat pieces, in order to optimize the contact area to sample volume ratio, so that a good internal equilibrium could be achieved. Typical masses of the samples employed were $m = 4-10$ mg for DSC and $m \approx 30$ mg for low-temperature specific-heat measurements. Brillouin measurements were performed using plan-parallel polished amber plates typically less than 0.5 mm thick.

After being completely measured and characterized, some pristine samples of amber were either partially or totally rejuvenated, by means of isothermal annealing processes near to or above $T_g$, respectively, followed by cooling at 1 K/min.

## 2.2 Differential Scanning Calorimetry (DSC)

Calorimetric characterization of amber through the different stages of rejuvenation (i.e. gradually erasing its thermodynamic history) was performed using a commercial DSC Q100 from TA Instruments. The technique employed was Temperature-Modulated Differential Scanning Calorimetry (TM-DSC), which allows to independently determine thermodynamic and kinetic stability via reversing and non-reversing contributions to the heat capacity, as well as the total heat capacity. Heating and cooling rates employed were typically ±1 K/min, and modulating signals ±0.5 K every 80 s.

In the case of Spanish amber, more accurate absolute values around 1% and higher reproducibility of the specific heat were possible by paying special attention to several experimental details: (i) the standard aluminum pans (mass $m \approx 20$ mg) used where chosen to deviate less than 0.1 mg from the reference; (ii) the sample mass was maximized to be approximately 10 mg; (iii) the sample was milled down to a homogeneous grain size of ≈ 50 µm, using an agate mortar; (iv) calibration of the specific-heat curves was done by measuring standard sapphire in the very same conditions as amber, and directly comparing it to the theoretical values, what provides a temperature-dependent correction factor. Accuracy of temperature data was estimated to be better than 0.1 K.

*2.3 Acoustic and elastic measurements*

In order to independently determine the Debye contribution to the low-temperature specific heat of the different amber samples, their sound velocity and mass density were obtained. The longitudinal sound velocity was measured in the temperature range 80 K < $T$ < 300 K using High Resolution Brillouin Spectroscopy (HRBS), with excitation wavelength $\lambda_0$ = 514.5 nm. Polished plan-parallel slabs of amber, less than 0.5 mm thick, were employed. Both backscattering (180°) and right-angle (90A) geometries were simultaneously used, the former implying a refractive-index dependent acoustic wave vector and the latter being independent of it. Given the high background signal introduced by the luminescence of the samples at these wavelenghts, the transverse sound velocities $v_T$ could only be measured at room temperature. Then, the zero-temperature values $v_T(0)$ and hence the Debye sound velocities $v_D$ were approximately obtained by means of the generalized Cauchy equation [24, 25, 29].

The density of the amber samples at room temperature was measured by employing the Archimedes method with a Mettler Toledo AB 265-S balance, using distilled water as a fluid. Linear extrapolation of the mass densities to zero temperature were performed by using the Lorenz-Lorentz relation between mass density $\rho(T)$ and refractive index $n(T)$ for a transparent medium, from our HRBS measurements in the range 80 K < $T$ < 300 K [24, 25].

*2.4 Low-temperature specific heat*

The low-temperature specific heat was measured by means of two complementary methods of thermal relaxation calorimetry [30], in the temperature range between 0.07 K and 30 K. Specific-heat measurements in the range 1.8K<$T$<30K were performed in a double-chamber insert, placed in a $^4$He cryostat, wheras measurements in the lowest temperature range 0.07K<$T$<3K were performed in a dilution refrigerator Oxford Instruments MX400. Calorimetric cells consisted of a sapphire disc, on which a small calibrated thermometer (either Cernox or RuO$_2$) and a resistive chip acting as a Joule heater were glued diametrically opposed, using cryogenic varnish GE7301. The sapphire substrate was suspended from a copper ring acting as thermally-controlled sink. The main thermal contact between the calorimetric cell and the thermal sink was a thin metallic wire through which heat is released. The heat capacity of the empty addenda was independently measured in each case and subtracted from the total data points. Excellent agreement was found between experimental data from both experimental setups in the overlapping temperature range. More details can be found in Refs. [24, 25].

## 3 Experimental results

First of all, it was deemed crucial to perform DSC experiments to confirm the presumed hyperaging and corresponding thermodynamic stabilization of the pristine samples of amber. As can be seen in Fig. 2(a) for Spanish amber, a huge endothermic peak was always observed during the first upscan, thereby confirming the strong stabilization occurred after (110 million years) *hyperaging*. The two subsequent (indistinguishable) curves correspond to the rejuvenated glass, after having been brought above the glass-transition temperature and cooled back at 1 K/min, and showed always good reproducibility. The same qualitative features were observed for the Dominican amber (see Fig. 1 of Ref. [24]) with $T_g \approx$ 380 K.

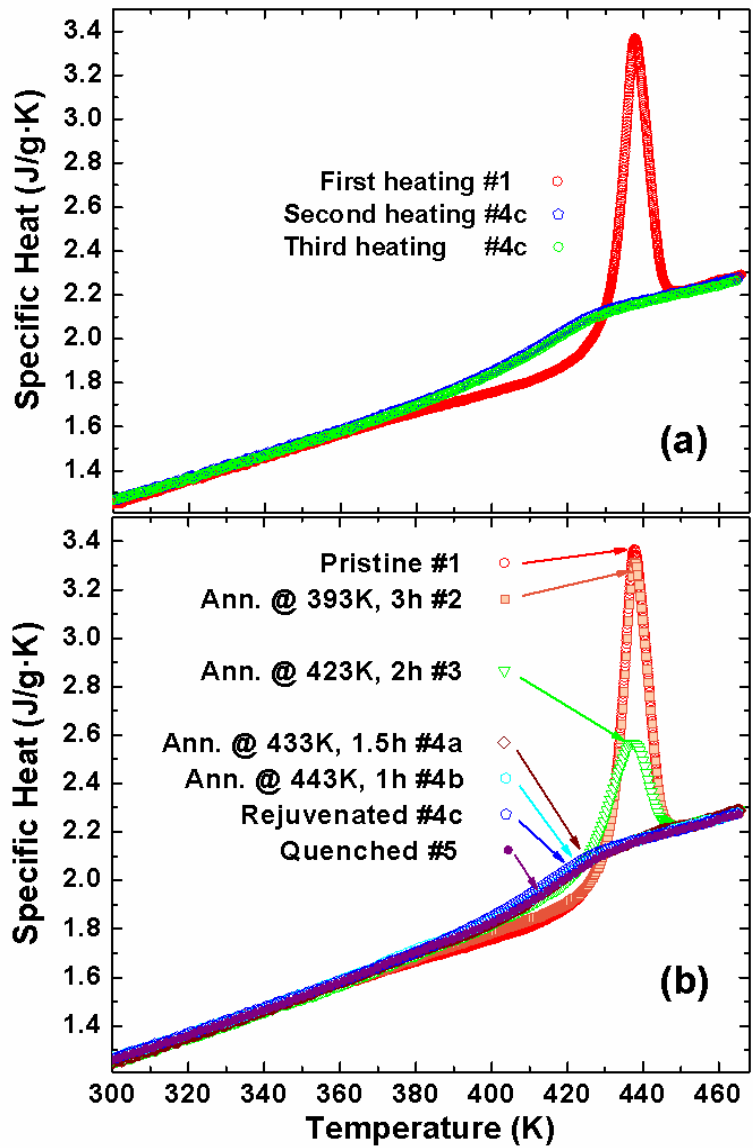

**Figure 2**. Specific-heat curves from DSC measurements in Spanish amber, systematically performed in all cases at rates ±1 K/min and modulating signal ±0.5 K every 80 s. (a) Typically obtained curves on a pristine amber sample corresponding to the first three consecutive heating runs. The first heating always shows a huge endothermic signal corresponding to the strong stabilization occurred after (110 million years) *hyperaging*. The two subsequent (indistinguishable) curves correspond to the rejuvenated glass, after heating it above the glass-transition temperature. (b) Specific-heat of Spanish amber after different isothermal treatments to the pristine sample near the glass transition (see legend). Only the first upscan for each sample is presented. The devitrification temperature of the pristine sample is located at $T_g^*$=438 K, and then decreases with decreasing stability (rejuvenation). The aging signal is clearly seen as a huge endothermic peak at the glass transition, which is maximal in the case of the pristine sample #1.

To study the evolution of amber samples after different thermal histories, we have determined their calorimetric glass-transition (strictly speaking, *devitrification*) temperature $T_g^*$ by the inflection point of the reversing $C_p$ jump from TM-DSC. Interestingly, the obtained $T_g^*$ = 438 K for the pristine amber is well above the genuine glass transition temperature $T_g$ = 423 K obtained for the rejuvenated sample, or alternatively from the second or third heating runs for any sample, when the cooling and heating rates are canonically the same. This unusual *increase* of the calorimetric $T_g^*$ for the stabilized amber compared to the canonical glass (see Table 1) is also observed in the case of ultrastable glasses obtained from physical vapor deposition [10, 31], and has been ascribed to a high *kinetic* stability, indicating that much higher temperatures are needed to dislodge the molecules from their glassy configurations.

Fig. 2(b) shows specific-heat curves of Spanish amber after different thermal treatments (only the first upscan for each sample is presented). As said above, the devitrification temperature of the pristine sample is located at $T_g^*$=438 K, and then decreases with decreasing stability (i.e. rejuvenation). Rejuvenation of amber was done stepwise (see Table 1) by performing different isothermal treatments to the pristine sample near the glass transition. The aging signal is clearly seen as a huge endothermic peak at the glass transition, which is maximal in the case of the pristine sample #1, thereby confirming that the geological aging of amber had indeed extraordinarily decreased its energy level in the PEL.

| Thermal history | # sample | $T_g^*$ (K) | $T_f$ (K) | $(T_f - T_f^0)/T_f^0$ |
|---|---|---|---|---|
| Pristine (hyperaged) | 1 | 438 | 384 | −7.7% |
| 3 h @ 393 K | 2 | 438 | 385 | −7.5% |
| 2 h @ 423 K | 3 | 436 | 391 | −6.0% |
| 1.5 h @ 433 K | 4a | 423 | 413 | −0.72% |
| 1 h @ 443 K | 4b | 423 | 415 | −0.24% |
| Rejuvenated (>460 K) | 4c | 423 | 416 | 0% |
| Quenched | 5 | 423 | 417 | +0.24% |
| Rejuvenated & annealed 2 h @ 423 K | 6 | 424 | 416 | 0% |

**Table 1**. Calorimetric data for Spanish amber: Glass-transition (devitrification) temperatures $T_g^*$ and fictive temperatures $T_f$ (see text for definitions) obtained after the different thermal histories applied to the studied samples. $(T_f - T_f^0)/T_f^0$ displays the relative decrease of the fictive temperature $T_f$ in relation to the canonical reference glass obtained after fully rejuvenation and cooling at 1 K/min (#4c).

Furthermore, the enthalpy as a function of temperature was obtained by direct integration from the calorimetric curves (see Fig. 2(b) of [25]). It is traditional and useful to determine the so-called *fictive* temperature $T_f$, defined as the temperature at which the nonequilibrium (glass) state and its equilibrium (supercooled liquid) state would have the same enthalpy [32]. The obtained values of $T_f$ for the pristine, partially rejuvenated and fully rejuvenated samples of Spanish amber are displayed in Table 1. The observed extraordinary

*decrease* $\Delta T_f = -32$ K (*thermodynamic* stability) for the pristine amber ($T_f$) compared to the rejuvenated glass ($T_f^0$) is similar or even superior to the effects seen in some ultrastable thin films of organic glasses [10, 31, 33, 34]. Such an extraordinary reduction ~8% of the fictive temperature due to the extremely long aging of amber can only be the consequence of extremely prolonged sub-sub-$T_g$ structural relaxations [35]. Let us remark that the fact of $T_g^* \neq T_f$ even for the canonical rejuvenated glass is mainly due to the specific calorimetric method used to determine $T_g^*$, i.e. the inflection point of the reversing $C_p$ jump from TM-DSC. For that reason, we prefer here to display in Table 1 the relative variation $(T_f - T_f^0)/T_f^0$. Interestingly, in the abovementioned paper by McKenna's group, they have also investigated by DSC amber from different sources, precisely finding that Dominican amber and our Spanish amber were the most stabilized ones, as indicated by unusually high $T_g - T_f$ values of 43.6 K and 41.4 K, respectively [28].

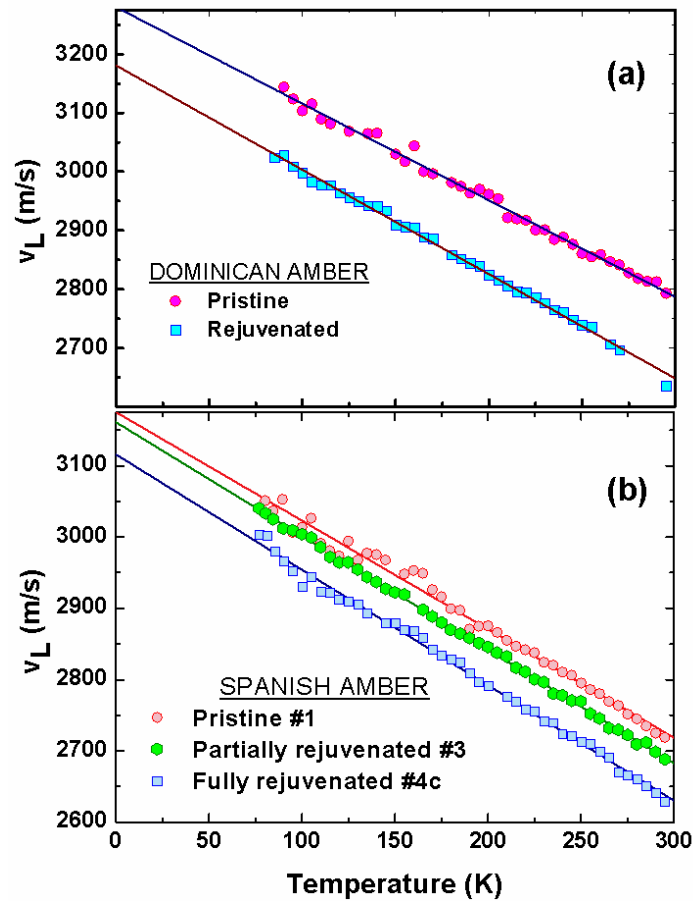

**Figure 3**. Temperature dependence of the longitudinal sound velocities measured by High Resolution Brillouin Spectroscopy in: (a) Dominican amber; (b) Spanish amber from El Soplao. In all cases, sound velocity gradually decreases with rejuvenation. Least-squares linear fits (solid lines) of the experimental data in the temperature range 80 K<$T$< 300 K are used to calculate the zero-temperature limit $v_L(0)$.

Another expected consequence of the dramatic stabilization process undergone by amber during its prolonged aging is *densification*, as well as an increase in the elastic constants. Since amber samples are heterogeneous and also present impurities, several pieces of the same sample before and after rejuvenation were always used, so that any possible spurious contribution could be minimized. Both mass density and sound velocity (see Fig. 3 and Table 2) were found to be around 2% (3%) higher in the pristine samples of Spanish (Dominican) amber than in their corresponding rejuvenated, conventional glasses. Hence, elastic moduli such as the longitudinal modulus, $M = c_{11} = \rho \cdot v_L^2$ have increased about 6% in the Spanish amber (more than 12% in the Dominican amber) due to the aging process, and the corresponding Debye cubic coefficients of the specific heat at low temperatures are 9% (14%) lower in the pristine samples of Spanish (Dominican) amber than in the fully-rejuvenated, canonical glasses.

|  | SAMPLE STATE | $\rho_{RT}$ (kg/m$^3$) | $\rho(0)$ (kg/m$^3$) | $v_L(0)$ (m/s) | $v_T(0)$ (m/s) | $v_D$ (m/s) | $c_D$ ($\mu$J g$^{-1}$ K$^{-4}$) | $T_{BP}$ (K) | $(C_P/T^3)_{BP}$ ($\mu$J g$^{-1}$ K$^{-4}$) |
|---|---|---|---|---|---|---|---|---|---|
| **Dominican amber** | Pristine (hyperaged) | 1058 | 1090 | 3280 | 1722 | 1926 | 15.5 | 3.6 | 62 |
|  | Fully rejuvenated (canonical glass) | 1028 | 1031 | 3180 | 1667 | 1865 | 18.0 | (3.6) | (56) |
| **Spanish amber** | Pristine (hyperaged) | 1045 | 1055 | 3175 | 1635 | 1831 | 18.9 | 3.4 | 51.9 |
|  | Partially rejuvenated | 1038 | 1049 | 3160 | 1625 | 1820 | 19.3 | 3.4 | 58.1 |
|  | Fully rejuvenated (canonical glass) | 1024 | 1035 | 3115 | 1596 | 1788 | 20.7 | 3.4 | 66.4 |

**Table 2**. Low-temperature elastic and specific-heat data for either Dominican amber or Spanish amber: Measured mass density at room temperature $\rho_{RT}$ and zero-temperature extrapolated $\rho(0)$, zero-temperature extrapolated longitudinal $v_L(0)$ and transverse $v_T(0)$ sound velocities, average Debye velocity $v_D$ and correspondingly calculated cubic Debye coefficient $c_D$ for the specific heat, temperature of the "boson peak" $T_{BP}$, and height of the "boson peak" $(C_p/T^3)_{BP}$.

In Fig. 4, Debye-reduced specific heat $C_p/T^3$ data are plotted in a wide low-temperature range for both kinds of amber. In each case, the very same sample is firstly measured in its pristine, as-received state, and then is partially or totally rejuvenated, in order to cancel out any possible spurious contribution to the specific heat due to the heterogeneity of the samples. As an example of this, two different samples of Dominican amber were measured and shown in Fig. 4(a). Solid symbols correspond to the pristine and rejuvenated states of the sample reported in Ref. [25] and open symbols correspond to another sample of Dominican amber, shown here for the first time, in order to check the reproducibility of the somewhat unexpected behaviour of the boson peak. Despite the very sensitive character of $C_p/T^3$ plots, only a minor quantitative shift is observed between both samples. Therefore, the most important result obtained by us is that highly-stable amber glasses exhibit both boson peak (the broad maximum in $C_p/T^3$) and TLS (the upturn below 1 K) well above the measured Debye levels $c_D$ (indicated by dashed lines in the figure), as any other conventional glass.

The specific heat of glasses at the lowest temperatures, usually ascribed to the tunneling TLS, is better studied by plotting $C_p/T$ vs $T^2$ to emphasize the TLS linear term (the intercept with the ordinate axis). The slope of the linear fit is related to the cubic Debye contribution, though one should be careful since the lower-$T$ tail of the boson peak would provide a non-negligible contribution here [36] (expected to be $C_p \propto T^5$ within the Soft-Potential Model), attributed to non-propagating vibrations. In each type of amber, the TLS linear term remains invariable, within experimental error, when erasing the thermal history. The small differences in their specific heat observed above 1 K are due to the corresponding variation in their Debye contributions. We obtain TLS coefficients of 12±2 $\mu J/gK^2$ for Dominican amber and of 5.8±0.4 $\mu J/gK^2$ for Spanish amber, as indicated by solid lines in Fig. 5. Strictly speaking, the TLS contribution at very low temperatures for Spanish amber was found to be better described by a *quasilinear* power law $C_p \propto T^{1.27}$ [25].

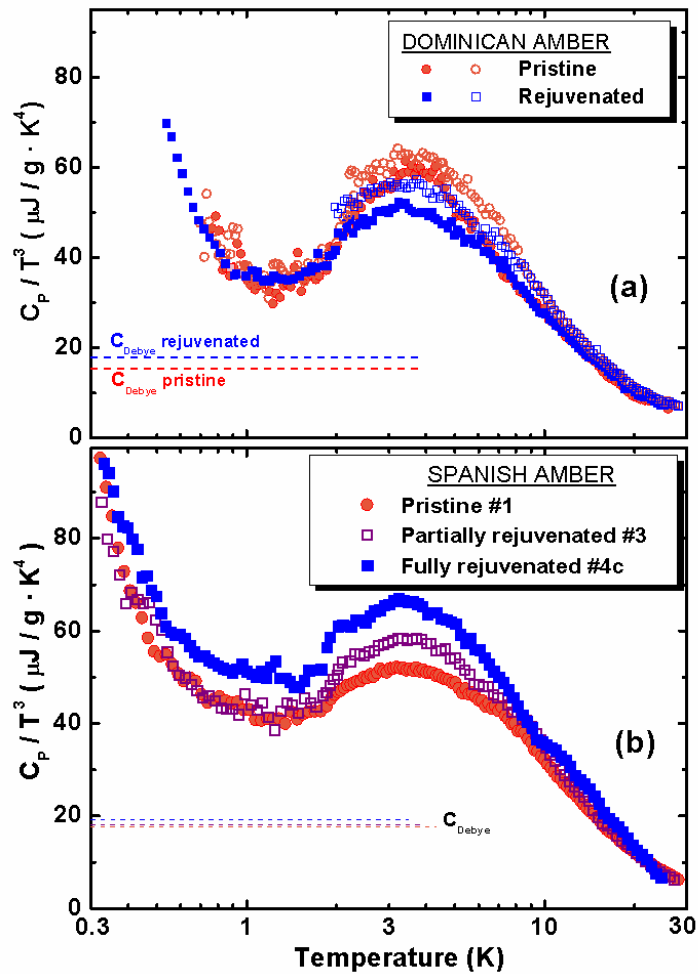

**Figure 4**. Debye-reduced specific heat $C_p/T^3$ at low temperatures for: (a) Dominican amber (solid and open symbols correspond to two different samples studied, see text); (b) Spanish amber. In each case, the very same sample is firstly measured in its pristine, as-received state, and then is partially or totally rejuvenated. The expected Debye contributions $c_D$ (see Table 2) are indicated by dashed lines.

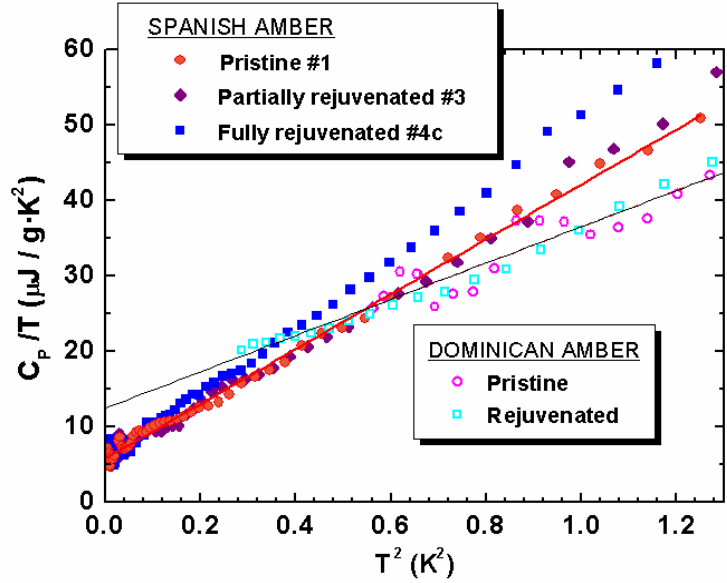

**Figure 5**. $C_p/T$ vs $T^2$ plot at the lowest temperatures, in order to emphasize the quasilinear contribution to the specific heat ascribed to TLS, for Dominican amber (open symbols) and for Spanish amber (solid symbols), as specified in the legend. Solid lines are linear fits assessing an invariable TLS-contribution of 12±2 μJ/gK$^2$ for Dominican amber (thin line) and of 5.8±0.2 μJ/gK$^2$ for Spanish amber (thick line).

## 4 Discussion

As said before, whether or not the low-temperature universal "anomalies" of glasses (dominated by low-energy excitations not present in crystals, i.e. tunnelling TLS and the boson peak) could be eventually suppressed by extremely long annealing or aging processes, and hence whether or not they are intrinsic properties of the glass state, has remained an unsolved question during the last forty years [17–23].

Let us focus first on the boson peak. From our experiments, it is clear that this vibrational excess over the Debye level persists in hyperaged glasses of amber. Notice that its position does not vary with rejuvenation, though a modest increase of its height is observed following the increase of the elastic Debye level in Spanish amber. Surprisingly, the $C_p/T^3$ boson peak is stronger in the 20-million-year old Dominican amber than in the rejuvenated one. This unexpected finding was observed to be reproducible by using different samples of Dominican amber. Nevertheless, an apparent residual curing or repolymerization of the samples was observed to occur around the glass transition temperature when rejuvenating those amber samples [24], hindering a reliable quantitative investigation in this case.

Despite some authors have tried to correlate the boson peak in glasses [37], and even in crystals [38], with transformations of the elastic continuum, such a Debye-scaling rule does not hold quantitatively in our case. The height of the $C_p/T^3$ boson peak in the hyperaged Spanish amber is 22% lower than that of the standard rejuvenated glass (see Table 2), whereas a Debye scaling [37] ($\propto \omega_D^{-3}$, with $\omega_D$ being the Debye frequency) would predict a reduction of only 7.4%. The situation in our Dominican amber is opposite and hence even worse. Although the hyperaged amber is significantly more dense and has higher sound velocity than the later rejuvenated glass, as expected, the specific heat around the sensitive boson peak is however slightly higher in the former than in the latter, as said above. Nonetheless, abovementioned apparent repolymerization processes occurring during rejuvenation cast doubts on any detailed quantitative discussion. In any case, the most relevant result found is that the boson peak featuring low-frequency dynamics of glasses is a robust property of noncrystalline solids that persists essentially unchanged in these very-long aged glasses, which are located very deep in the energy landscape, as thermodynamic characterization proves. The exact interrelation between the boson peak magnitude and the Debye coefficient, if any, cannot be deduced from these measurements.

Notwithstanding this, the best fingerprint of the universal glassy anomalies is likely the density of TLS, measured from the corresponding quasilinear contribution to the specific heat, since the influence of Debye-like lattice vibrations becomes less and less important below 1 K, and because these excitations have no counterpart in genuine crystalline solids. In this respect, our experimental results are conclusive [24, 25]: pristine and rejuvenated glasses from both kinds of amber exhibit the same specific heat below 1 K, within experimental error, especially if the small differences in their Debye contributions are discounted.

Therefore, the boson peak and, especially, the tunnelling TLS appear to be robust and intrinsic properties of glasses which remain "fossilized" in hundred-million-year stabilized glasses of amber. Furthermore, since our amber samples can be considered typical bulk polymer glasses, there is no reason not to extrapolate this conclusion to any standard glass.

It is however worth mentioning the apparently contradictory result previously found by us in physical-vapour deposited ultrastable glasses of indomethacin [10]. In brief, these ultrastable glasses did exhibit a full suppression (within experimental error) of the TLS contribution to the specific heat, whereas similar samples of the same substance lacking such ultrastability exhibited a typical glassy behaviour. Nevertheless, it was argued that this unexpected effect is a consequence of the very anisotropic and layered character of so-grown ultrastable glasses and not of their special thermodynamic state. In the same line of reasoning, it has also been shown recently that in some thin films of amorphous silicon (a particularly rigid amorphous network) the TLS can be removed by increasing the atomic density of the film, which depends on both the film thickness and growth temperature [39, 40].

Finally, we expect that amber will serve in the near future as an extremely enlightening model glass to study many other puzzles involved in the physics of the glass state. For instance, a very recent study by stress relaxation experiments as a function of temperature on 20-million-year-old Dominican amber, has shown that the instantaneous relaxation time dramatically deviates from the behaviour expected within the diverging-relaxation-time scenario [41].

## 5 Conclusion

Up to date, no study had been performed to investigate the possible influence that the dramatic increase in thermodynamic and kinetic stability occurring in geologically aged glasses, ideally approaching the "ideal glass" state with zero configurational entropy, could have on their universal low-temperature anomalies.

Our experiments on hyperaged glasses of amber from two different sources and hence chemical composition and thermal history, have undoubtedly demonstrated that the low-temperature "anomalous" properties of glasses (i.e. TLS and the boson peak) persist essentially unchanged in ideal-like glasses, subjected to a dramatic thermodynamic and kinetic stabilization. Therefore, they are robust and intrinsic properties of glasses.

In our view, amber can be an extremely useful benchmark for the glass research field, which hopefully could shed light on several of the puzzles and long-standing unsolved questions about the physics of glasses, including the glass transition.

**Acknowledgments**

The Laboratorio de Bajas Temperaturas (UAM) is an associated unit with the ICMM-CSIC. We acknowledge financial support by the Spanish Ministry of Economy and Competitiveness under project grants FIS2011-23488, MAT2012-37276-C03-01 and MAT2014-57866-REDT, and by the Autonomous Community of Madrid through program NANOFRONTMAG-CM (S2013/MIT-2850).